\documentstyle[aps,twocolumn,epsf,floats]{revtex}
\begin{document}
\draft
\preprint{MIC-DFM preprint}
\title{Magneto-Coulomb drag:
interplay of electron--electron interactions and Landau quantization}

\author{Martin Christian B\o nsager,$^{1,2}$
Karsten Flensberg,$^{1,3}$
Ben Yu-Kuang Hu,$^1$
and Antti-Pekka Jauho$^1$}

\address{$\mbox{}^1$ Mikroelektronik Centret,
Bygn. 345\o, Danmarks Tekniske Universitet,
DK-2800 Lyngby, Denmark}
\address{$\mbox{}^2$ Department of Physics,
Indiana University, Bloomington, Indiana 47405-4202}
\address{$\mbox{}^3$ Dansk Institut for Fundamental Metrologi,
Bygn. 307, Anker Engelunds Vej 1,
DK-2800 Lyngby, Denmark
\medskip\\
\date{\today}
\parbox{14cm}{\rm
We use the Kubo formalism to calculate the transresistivity $\rho_{21}$
for carriers in coupled quantum wells in a large perpendicular magnetic
field $B$.
We find that $\rho_{21}$ is enhanced by approximately 50--100 times over
that of the $B=0$ case in the interplateau regions of the integer
quantum Hall effect.
The presence of both electron--electron interactions and Landau
quantization results in (i) a twin-peaked structure of $\rho_{21}(B)$ in
the inter-plateau regions at low temperatures,
and, (ii)
for the chemical potential at the center of a Landau
level band, a peaked temperature dependence of $\rho_{21}(T)/T^2$.
\smallskip\\}}
\maketitle
\narrowtext

The combination of electron--electron (e-e) Coulomb interaction
and strong magnetic ($B$) field in two-dimensional electron gases
(2DEGs) has provided an exciting venue of research for both experimentalists
and theorists over the past few decades\cite{girvin}.
One well-known example of this is the fractional quantum Hall effect,
where the physics is determined by
the subtle interplay between the interactions and the
large density of states (DOS) caused by all the electrons being confined
to the lowest Landau level (LL).
Even in the integer quantum Hall effect, where e-e interactions
do not play such a crucial  role, they are thought to determine
some important factors such as the position of the edge currents\cite{chkl92}.
On the other hand, e-e interactions in parabolic confined systems
in a magnetic field surprisingly have {\em no}
effect on cyclotron resonance measurements
(due to generalized Kohn's theorem)\cite{kohn}.
Thus, phenomena involving inter-electron interactions
in a $B$-field often produce surprising and interesting results.

Recently, there have been many experiments on coupled
2DEGs 
which have probed
the effect of Coulomb interactions, both with and without
magnetic field.  Some experiments measured tunneling from one
well to the other\cite{murp95},
while in others the quantum wells were separated by a distance
at which interwell tunneling was negligible but interwell
Coulomb interactions were experimentally detectable\cite{exp_all}.
The latter ``drag'' experiments, so-called because one drives
a current in one layer and measures the consequence of the
frictional drag due to the interlayer interactions in the second layer,
provide a direct measure of the interwell Coulomb interaction.
Not surprisingly, physics centered around drag phenomena has
generated many theoretical
investigations\cite{dragtheo,jauh93,zhen93,shim94,kame95,flen95a,flen95b}.
In principle, drag experiments should provide a unique forum for exploring
the subtleties of the interplay of e-e interactions a magnetic field.
Thus far, however, only zero magnetic field data have been published.

In this Letter, we present a first-principles formulation of the
drag problem in a magnetic field, including effects
due to weak impurity scattering, starting from Kubo theory.
We then show results of an explicit numerical
calculation of the transresistivity $\rho_{21}^{xx}$
for short-ranged scatterers in the inter-plateau regions of the integer
quantum Hall regime.
We demonstrate that the aforementioned large DOS and screening due
to intralayer e-e interactions have profound effects on the $\rho_{21}$.


In principle, a drag experiment can be performed by imposing
a fixed electric field ${\bf E}_1$ on the ``drive'' layer
(henceforth called layer 1)
and measuring the current ${\bf J}_2$ dragged along in the
``response'' layer (called layer 2), placed a distance $d$ away.
The Kubo formalism allows one to compute
the transconductivity $\tensor\sigma_{ij}{\bf E}_j = {\bf J}_i$
($i=1,2$), which we can invert to obtain the transresistivity
$\tensor{\rho}_{ij} {\bf J}_j = {\bf E}_i$\cite{dragrate}.
For time-independent transport, to second
(i.e., lowest nonvanishing) order in the screened
interlayer interaction $W_{12}(q,\omega)$,
$\tensor\sigma_{21}$ is given by\cite{kame95,flen95a,uniform}
\begin{eqnarray}
\sigma_{21}^{\alpha\beta} &=&
\frac{e^2}{2\hbar^3} \int \frac{d{\bf q}}{(2\pi)^2}
\int_{-\infty}^\infty \frac{d\omega}{2\pi} |W_{12}(q,\omega)|^2
\left(-\frac{\partial n_B(\hbar\omega)}{\partial \omega}\right)\nonumber\\
&\times& \Delta_2^\alpha({\bf q},{\bf q};\omega+i0^+,\omega-i0^+)
\nonumber\\
&\times&\Delta_1^\beta(-{\bf q},-{\bf q};-\omega-i0^+,-\omega+i0^+),
\label{sigma21}
\end{eqnarray}
where $n_B$ is the Bose function and
${\bf \Delta}_l$ is the imaginary-time Fourier transform of
the thermal-averaged correlation function
$i \langle {\cal T}_\tau\; {\bf j}_l({\bf q}=0,\tau=0)
\rho_l({\bf q},\tau) \rho_l(-{\bf q},\tau') \rangle$\cite{kame95,flen95a}.
Screening is calculated using the random phase approximation
for electrons in a magnetic field with weak impurity
scattering\cite{ando-san}, where the density-response function
$\chi(q,\omega)$ is given diagrammatically by Fig.\ 1(a).
We assume throughout this paper that there are
like charges in both layers (generalization to unlike charges
is straightforward), and that spin-splitting is negligible\cite{folg95}.

We let $x-y$  be the
confinement plane for the electrons, ${\bf B} = B\hat z$,
and use the Landau gauge ${\bf A}  = (0,Bx,0)$.
${\bf \Delta}({\bf q},{\bf q},\omega\pm i0^+ ,\omega\mp i0^+)$
is a real, gauge invariant quantity.
Ignoring diagrams with crossed impurity
lines
\begin{figure}
\epsfxsize=7.5cm
\epsfbox{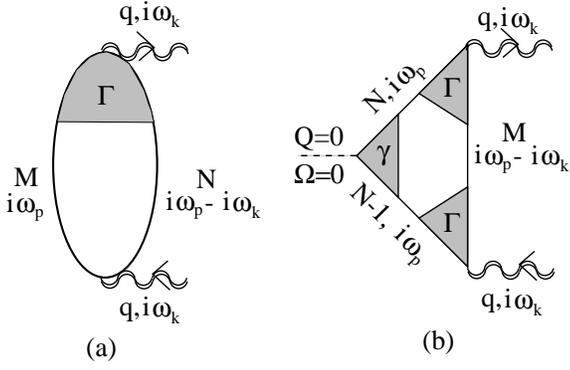}
\vspace{0.5cm}
\caption{
(a) The diagram corresponding to the density response function $\chi$.
(b) The triangle diagram contribution to the 3-body response function
${\bf \Delta}$.  The $\Gamma$ and $\gamma$ are the charge and current vertices,
respectively, and the labels $M$ and $N$ denote the LL's.
}
\label{feynman}
\end{figure}
\noindent
(which are negligible in the weak-scattering and high $B$
field limit), ${\bf \Delta}$ is shown diagrammatically in Fig. 1(b)
and can be written down in terms of the single-particle Green functions
$G$, and charge and current vertices.
For 2DEGs 
in a quantizing
magnetic field it is crucial to include the impurity effects
from the outset to avoid non-physical results.   We do this
within the self-consistent Born approximation (SCBA)\cite{ando-san}.
Then, the Green functions and the self-energy $\Sigma$
depend only on the LL index\cite{ando-san},
$G(N,i\omega_n) =
\left[i\omega_n - \varepsilon_N -
\Sigma(N,i\omega_n)\right]^{-1}$.
In calculating ${\bf \Delta}$, we include all
all ladder-type diagrams as required by the Ward identity.

The general expression for ${\bf \Delta}$ for arbitrary scattering
is complicated, but many simplifications occur
in the weak-scattering limit $\omega_c\tau \gg 1$, where $\omega_c =
eB/m$ is the cyclotron frequency and $\tau$ is the Born approximation
scattering time. In particular, it is possible to link ${\bf \Delta}$
with the $\chi({\bf q},\omega)$ shown in Fig. 1(a)\cite{tritobub}.
We find
\begin{eqnarray}
{\bf \Delta}({\bf q},{\bf q};\omega\pm i0^+,\omega\mp i0^+)=\ \ \ \
&&\nonumber\\
\pm 2\hbar^2 e^{-1}
{\bf q}\times {\bf B}\ \frac{\mbox{Im}[\chi({\bf q},\omega)]}{B^2}
&&+ O\Bigl((\omega_c\tau)^{-1}\Bigr).
\label{delta}
\end{eqnarray}
It is worth emphasizing that for $\omega_c\tau \gg 1$,
the relationship between ${\bf \Delta}$
and Im$[\chi]$ holds for {\sl arbitrary} impurity scattering potential
$U({\bf q})$, whereas ${\bf\Delta}(B=0)$ is related to Im$[\chi]$
only for $q$-independent $U$\cite{flen95a,flen95b,physexpl} .

\begin{figure}
\epsfxsize=7.5cm
\epsfbox{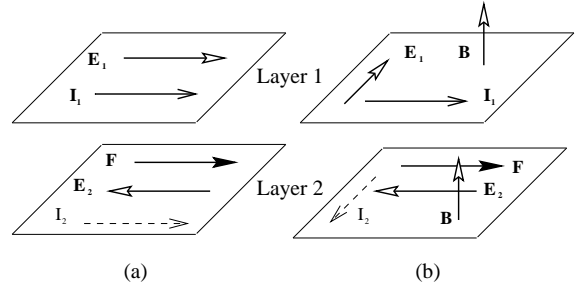}
\vspace{0.5cm}
\caption{
Schematic illustration of the sign of the diagonal elements of
the $\tensor{\sigma}_{21}$ and $\tensor{\rho}_{21}$
at (a) B=0 and (b) strong B field (charges assumed positive).
${\bf I}_i$ are the currents, ${\bf E}_i$ are electric fields, and
${\bf F}$ is the average net force transmitted from layer 1 to 2.
For a transresistivity (transconductivity) measurement,
${\bf I}_2 = 0$ (${\bf E}_2 = 0$) and
${\bf E}_2$ (${\bf I}_2$) is measured.  While
$\sigma^{\alpha\alpha}_{21}$ changes sign going from (a) to (b),
$\rho^{\alpha\alpha}_{21}$ does not.
For transresistivity measurements, ${\bf E_2} \| {\bf F}$
because ${\bf F}-e({\bf E_2}+ \langle{\bf v_2}\rangle\times{\bf B})
= 0$, and $\langle{\bf v_2}\rangle=0$.
Hence, under conditions where ${\bf I}_1 \| {\bf F}$,
which is the case when when there is inversion symmetry (e.g., when $B=0$),
or when the electron distribution is a drifted Fermi-Dirac
(e.g., when $\omega_c\tau \gg 1$), there is no Hall transresistivity.
}
\label{physical}
\end{figure}

A brief discussion of the $\pm$ sign occuring in the
high field limit of Eq. (\ref{delta}) is appropriate.
The Onsager relation and the vector nature of ${\bf\Delta}$ imply that
it must have form\cite{bons96}
${\bf\Delta}({\bf q},{\bf q};\omega\pm i0^+,\omega\mp i0^-, {\bf B}) =
{\bf q}\, u(q,B,\omega) \pm ({\bf q}\times{\bf B})\, v(q,B,\omega)$.
The ${\bf q}\, u$ term dominates for small $B$, while the
$({\bf q}\times{\bf B})\, v$ term dominates for $\omega_c\tau \gg 1$,
which is consistent with Eq. (\ref{delta}).
The form of ${\bf \Delta}$ implies from Eq. (\ref{sigma21})
that as $B$ is increased from $0$,
$\sigma^{xx}_{21}$ changes sign at some point.
Does this mean a change in sign of an experimentally measured quantity?
If the measured quantity is the transresistivity
${\rho}_{21}^{xx}$,
as is usually the case\cite{exp_all}, the answer is no,
for the following reason.
In terms of $\tensor{\sigma}_{ij}$,
$\tensor{\rho}_{21} = [-\tensor{\sigma}_{11}\tensor{\sigma}_{21}^{-1}
\tensor{\sigma}_{22} + \tensor{\sigma}_{12}]^{-1}\approx
-\tensor{\sigma}_{22}^{-1}\tensor{\sigma}_{21}\tensor{\sigma}_{11}^{-1}$
(since $|\tensor{\sigma}_{ii}|\gg |\tensor{\sigma}_{21}|$).
For $B=0$, $\tensor{\sigma}_{ii}$ are diagonal and
$\rho_{21}^{xx}(B\!=\!0) =
-\sigma_{21}^{xx}/(\sigma_{11}^{xx}\sigma^{xx}_{22})$;
i.e., $\rho_{21}^{xx}$ and $\sigma_{21}^{xx}$ have {\sl opposite} signs.
In contrast, for quantum Hall systems, $|\sigma^{xy}_{ii}| \gg
\sigma^{xx}_{ii}$ and hence $\rho^{xx}_{21}(\omega_c\tau\!\gg\! 1)
\approx -\sigma_{21}^{yy}/ (\sigma_{11}^{xy}\sigma_{22}^{yx})$.
Since $\sigma_{11}^{xy}\sigma_{22}^{yx} < 0$, this implies that
$\sigma_{21}^{xx}(\omega_c\tau\!\gg\! 1)$
and $\rho_{21}^{xx}(\omega_c\tau\!\gg\! 1)$ have the {\sl same} sign.
Therefore, even though $\sigma_{21}^{xx}$ changes sign
as $B$ increases, the experimentally
relevant quantity $\rho^{xx}_{21}$ is negative in both cases.
A physical explanation of this result is illustrated
in Fig. \ref{physical}.


Eqs. (\ref{sigma21}) and (\ref{delta}) form the basis of our
numerical calculations.
We obtain $\chi(q,\omega)$ by solving the
appropriate vertex equation, and
perform the integrals in (\ref{sigma21}) to obtain $\tensor\sigma_{21}$,
and consequently $\tensor\rho_{21}$.
We discuss the technical details
\begin{figure}
\epsfxsize=7.5cm
\epsfbox{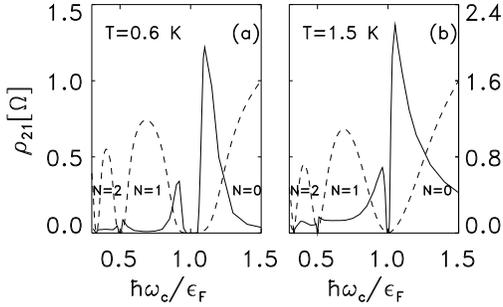}
\vspace{0.5cm}
\caption{
Transresistivity $\rho_{21}$ (solid lines) and
and the thermally averaged DOS $g = \partial n/\partial \mu$ (dashed,
in arbitrary units) for (a) $T=0.6\,{\rm K}$, and (b) $T=1.5\,{\rm K}$
as a function of magnetic field
in GaAs for density $n=1.5\times 10^{11}\,{\rm cm}^{-2}$
($E_F/k_B\approx 60\,{\rm K}$), well separation $d=350\,{\rm\AA}$
and zero well widths. $N$ is the LL index, and $\hbar\omega_c = \epsilon_F$
corresponds to $B = 3.1\,{\rm T}$. While the $g(B)$
peaks in the middle of the Landau level, the interlayer coupling
is weakest there (due to large screening), pushing the peaks in
$\rho_{21}$ towards the edges of the Landau level bands.
}
\label{Bdepn}
\end{figure}
elsewhere\cite{bons96},
and focus here on the qualitative features and
some of the numerical results.

Without impurities, all electrons in a particular LL are degenerate.
When the scattering is weak (so that inter-LL coupling can be ignored)
within the SCBA with short-ranged scatterers
(i.e., scattering interaction range $\ll\ell_B/\sqrt{2N+1}$),
$\Sigma(N,i\omega_n)$ is $N$-independent,
and the electrons in a LL are distributed in
bands where the DOS is semi-elliptical\cite{ando-san}.
The DOS at the center of the LL is approximately
$\sqrt{2\omega_c\tau\pi^{-1}}\, g_0$, where
$g_0 = m\pi^{-1}\hbar^{-2}$ is the 2-dimensional zero magnetic field DOS.
Thus, when $\omega_c\tau \gg 1$ (achieved in clean GaAs samples at
fields under a Tesla) the DOS is greatly enhanced over the $B=0$ value.
The low-temperature transresistivity for fixed $T$ is to a first
approximation directly proportional to the product of the
thermally averaged DOS of both layers, $g_1(B)\, g_2(B)$,
around the chemical potential $\mu$
(since the more phase-space there
is available for scattering around the Fermi surface,
the larger the probability for interlayer momentum transfer).
Hence, one might expect that
(1) $|\rho_{21}^{xx}(\omega_c\tau \gg 1)| \gg  |\rho_{21}^{xx}(B=0)|$, and
(2) $\rho_{21}^{xx}(B)$ would more or less simply reflect the
shape of $g_1(B)\, g_2(B)$.

Fig.\ 3 shows the results of a calculation for $\rho_{21}^{xx}(B)$
for two identical layers at fixed densities.  For comparison, we also
show $|{\rm Re}[\chi(q\rightarrow 0,\omega=0)]|= \partial n/\partial \mu
\equiv g$\cite{chiDOS}.
As expected, the $\rho_{21}^{xx}$
is very large; approximately 50 -- 100 times larger than at $B=0$\cite{hill}.
Also, the $\rho_{21}$ is largest when $\mu$ is in the bands of
extended states, and suppressed when it is in between the extended
bands\cite{shim94,local}.
However, the shape of $\rho_{21}(B)$ is markedly different from
$g^2(B)$.  Relative to $g^2(B)$, there is an
enhancement in $\rho_{21}(B)$ at the edges of the broadened LL
and suppression at the center.
\begin{figure}
\epsfxsize=7.5cm
\epsfbox{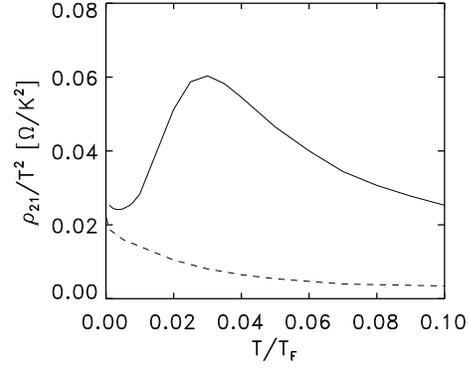}
\vspace{0.5cm}
\caption{
Transresistivity as a function of temperature, for $B=2.1\,$T (equivalent
to a half-filled $N=1$ LL), using dynamical screening (solid line) and static
screening (dashed line).
Other parameters are as in Fig.\ \protect\ref{Bdepn}.
The peak in the solid curve is
caused by the decrease in the screening ability of the electron
gases for frequencies larger than $\ell_B^2q^2/\tau$, whereas
the static screening curve falls monotonically due to the decrease
in density of states, which reduces the number of electrons
participating in the drag.
}
\label{Tdepn}
\end{figure}

This effect originates from the screening
properties of the system.  Recall that $\rho_{21}$ also depends
on interlayer coupling, which is given by the screened interlayer
interaction $W_{21}$.   Roughly,
$\rho_{21}$ is proportional to $g_1\,g_2 \times
|W_{21}|^2$.
For 2DEGs, 
the range of the screened interaction varies inversely
with $g$\cite{ando-san}.
Therefore, increasing $g(B)$ weakens the interlayer
coupling, implying that the terms
$g_1\, g_2 $ and $|W_{21}|^2 $ tend to work
in opposition.  This results in the following scenario when $B$ is changed.
When $\mu$ lies in the region of localized
states below a LL band, $\rho_{21}$ is very small because
very few electrons have sufficient energy to be
excited into extended states where they contribute to
the drag\cite{platvalid}.
As $B$ is increased so that $\mu$ moves into the LL band, the density of
extended states increases, while the interlayer interaction is strong
due to weak screening, resulting in a sharp rise in $\rho_{21}$.
However, as the $B$ field is further increased
so that $\mu$ moves closer towards the center
of the LL and the DOS further increases, the screening becomes
so effective that it more than compensates for the increase in DOS,
leading to a reduction in $\rho_{21}$.   This competition of DOS and
screening produces the unique shape of $\rho_{21}(B)$.

We also find interesting behavior in $\rho_{21}$ when
$B$ is kept constant and the temperature $T$ is changed.
We concentrate on the $T$-dependence for $\mu$ in the
middle of a LL band.  If the DOS were constant
and the interaction were $\omega$-independent, the scaled
transresistance $\rho_{21}/T^2$ would be $T$-independent\cite{jauh93}.
Fig. 4, however, shows that $\rho_{21}/T^2$ rises at
$k_B T/E_F \approx 0.01$ and peaks at $k_B T/E_F \approx 0.03$.
This effect is also attributable to screening, as the following
shows.

An examination of $\chi(q,\omega)$ shows a sharp fall in the
screening as a function of $\omega$ at frequency scales consistent with
the temperature at which $\rho_{21}/T^2$ rises\cite{bons96}.
In a large $B$-field, the diffusive form of the polarizability
$\chi_{\rm diff}(q,\omega) \sim -D q^2/(D q^2 - i\omega)$ is valid
for $q\ell_B \ll 1$ and $\omega\tau\ll 1$.
Since electrons hop to adjacent orbits a distance $\ell_B$ away
on average every $\tau$, the diffusion constant $D\approx \ell_B^2/\tau$.
For a given $q$, the maximum of
${\rm Im}[\chi_{\rm diff}(q,\omega)]$ occurs at $\omega = Dq^2$.
We take $d^{-1}$ to be a typical $q$ value,
since $W(q,\omega)$ cuts off the integral in Eq.\
(\ref{sigma21}) at this wavevector.  This gives a peak in ${\rm
Im}[\chi_{\rm diff}(d^{-1},\omega)]$ at
$\omega\sim\ell_B^2/(d^2\tau) \approx 0.01 E_F$ for the parameters
used in Fig. 4. For $\omega$ higher than this,
both Re$[\chi]$ and Im$[\chi]$ fall off rapidly, decreasing the screening.
Thus, as the temperature rises above $0.01 E_F$, the reduction in
screening increases the effective interaction
and hence the scaled $\rho_{21}$.  Note that this temperature dependence
is a direct consequence of the diffusive nature of the system, and it can
be seen at experimentally viable temperatures,
in stark contrast to $B=0$, where the diffusive nature
of the system only manifests itself in $\rho_{21}$
at unattainably low temperatures\cite{zhen93}.
This difference comes about because $\chi_{\rm diff}$ at $B=0$
is valid only for $q \ll (v_F\tau)^{-1}$, which is a much smaller
$q$-region of validity than for large $B$, since $v_F\tau \gg \ell_B$.

Summarizing, we have presented a microscopic calculation of transresistivity
for Coulomb coupled quantum  wells in strong magnetic fields.  Both
magnetic field and temperature dependence of the transresistivity are
clearly distinct from normal longitudinal magnetoresistivity; the
differences arise from an intricate interplay between Landau quantization,
interparticle interaction and diffusion effects.

We thank A. H. MacDonald and A. Wacker for useful comments, and
Nicholas Hill for sharing his unpublished experimental results
with us.  KF was partially supported by the Carlsberg Foundation.


\begin{thebibliography}{10}
\bibitem{dragrate}
The transresistivity is related to the drag rate $\tau_{D}^{-1}$, which is
  defined and used in several papers in Refs. \onlinecite{exp_all} and
  \onlinecite{dragtheo}, by $\rho_{21}^{\alpha\alpha} = m_1/(n_1e^2 \tau_{D})$.
\bibitem{uniform}
This form is for a uniform system, and ignores edge effects. We expect this
  form to be valid in this paper, since the significant results we present are
  for the interplateau regions, where the eigenstates are extended and
  bulklike.
\bibitem{tritobub}
The 3 vertex corrections and 3 $G$'s in ${\bf\Delta}$ are reduced to 1 vertex
  and 2 $G$'s in $\chi$ as follows. (i) {\sl Vertices:} When $\omega_c\tau\gg
  1$, vertex corrections are significant only when the incoming LL indices are
  equal. The $\gamma$ can be neglected because the LL indices of the incoming
  $G$'s differ by 1, and of the remaining $\Gamma$'s, only one can have equal
  incoming indices. (ii) {\sl Green functions:} $G(N,\omega)G(N-1,\omega)
  \approx [G(N,\omega)-G(N-1,\omega)]/\omega_c + O((\omega_c\tau)^{-1})$,
  reducing 3 $G$'s to 2. This creates ``mismatches'' between the LL indices of
  the bare vertex functions $f_{M,N}({\bf q})$ 
  and the $G$'s (i.e., it produces $G(M)G(N)f_{M,N\pm 1}$); the matches
  are restored and a factor of
  ${\bf q}$ created with the identity $\sqrt{N+1}f_{N+1,M}({\bf q}) - \sqrt{M}
  f_{N,M-1}({\bf q}) \equiv - \ell_B(iq_x + q_y){\sqrt 2}^{-1} f_{N,M}({\bf
  q})$, where $\ell_B=(\hbar/eB)^{1/2}$ is the magnetic length.
\bibitem{physexpl}
This can be understood physically by considering the semi-classical electron
  distribution function. The relationship $\Delta\propto {\rm Im}[\chi]$
  results when the distribution function in the linear current response regime
  is a drifted Fermi-Dirac, which occurs both for constant $U$ at $B=0$ and for
  arbitrary $U(q)$ at large $B$-fields.
\bibitem{platvalid}
While we cannot say anything quantitative about the quantum hall plateau regime
  because our model excludes some of the pertinent physics here, we expect
  $\rho_{21}$ to be small because there is an energy gap to excitations.
\end{thebibliography}
\end{document}